  \documentclass[a4paper, 11pt]{article}
\usepackage{latexsym}
\usepackage{graphicx}
\usepackage{epsfig}

\begin{document}

\begin{titlepage}

\begin{center}

{\Large\bf\sc Virus Structure:}

\vskip .2cm

{\Large\bf\sc From Crick and Watson to a new conjecture$^*$}

\vskip 1cm

{\Large Alfredo Iorio$^{a,b}$\footnote{E-mail: iorio@ipnp.troja.mff.cuni.cz} and
Siddhartha Sen$^{c,d}$\footnote{E-mail: tcss@mahendra.iacs.res.in}}

\vskip 1cm

$^{a}$ Institute of Particle and Nuclear Physics, Charles
University of Prague  \\
V Holesovickach 2, 18200 Prague 8 - Czech Republic\\
$^b$ Department of Physics ``E.R.Caianiello'' University of Salerno and INFN \\
Via Allende, 84081 Baronissi (SA) - Italy \\
$^c$ School of Mathematical Science, University College Dublin \\
Belfield, Dublin 4 - Ireland \\
$^d$ Indian Association for the Cultivation of Science \\
Jadavpur, Calcutta 700032 - India

\end{center}

\vskip 1cm

\begin{abstract}
We conjecture that certain patterns (scars), theoretically and numerically predicted to be formed by electrons arranged on a sphere to minimize the repulsive Coulomb potential (the Thomson problem) and experimentally found in spherical crystals formed by self-assembled polystyrene beads (an instance of the {\it generalized} Thomson problem), could be relevant to extend the classic Caspar and Klug construction for icosahedrally-shaped virus capsids. The main idea is that scars could be produced at an intermediate stage of the assembly of the virus capsids and the release of the bending energy present in scars into stretching energy could allow for a variety of non-spherical capsids' shapes. The conjecture can be tested in experiments on the assembly of artificial protein-cages where these scars should appear.
\end{abstract}

\vfill

\noindent $^*$ Invited talk by A.I. at the Fourth International Summer School and Workshop on Nuclear Physics Methods
and Accelerators in Biology and Medicine - Prague, 8-19 July 2007.



\end{titlepage}


\section{Virus Structure}

\paragraph{General Considerations} Viruses are small pieces of genetic material (DNA or RNA) that can
efficiently encode few small identical proteins that then assemble themselves\footnote{Sometimes, for large viruses, the assembly is done with the help of other proteins encoded to this end by the genetic material. The environment also plays an important role.} to form ``cages'' around the genetic material \cite{general}. These cages are called {\it capsids} and their shape is the main concern of this paper.

Capsids are essential for protecting the genetic material and contribute to identifying cells suitable for the duplication of the genetic
material\footnote{Viruses are not able to duplicate without the help of the host cell, that is why their living nature is debatable.}. Understanding the way proteins arrange to form these very resistent capsids is important: if we could find a way to undo these constructions we would be able to destroy viruses. There are three classes of capsid's shapes \cite{general}: {\it helical} (the proteins spiralize counter-clockwise
around the genetic material), {\it icosahedral or simple} (the proteins arrange in morphological units of 5 and 6 following precise geometrical
and topological prescriptions, as we shall soon explain), {\it complex} (sphero-cylindrical, conical, tubular or even more complicated shapes, i.e. without a precise resemblance to any particular regular polyhedron). There are also {\it polymorphic viruses} that change their shape, e.g., from icosahedral to tubular and {\it enveloped viruses} that, in addition to the protein-capsid, also have an outer lipid bilayer (the viral envelope) taken by the host cell membranes.

\paragraph{Icosahedral Viruses} In 1956 Crick and Watson \cite{crickwatson} proposed that small viruses have capsids with the identical
proteins (or {\it subunits} or {\it structural units}) arranged into morphological units called {\it capsomers} with the shape of hexagons and pentagons, called {\it hexamers} and {\it pentamers}, respectively. These capsomers form polyhedrons that go under the name of {\it icosadeltahedrons}, with a fixed number, 12, of pentamers and a variable number of hexamers. Following Crick and Watson's seminal idea, Caspar and Klug (CK) \cite{casparklug} later extended the class of viruses to which this construction applies to what they called ``simple viruses'', i.e. still roughly spherical but not necessarily small viruses. The CK model for icosadeltahedral capsids is nowadays universally accepted by virologists \cite{general}, \cite{viper}.

The fact that exactly 12 pentamers are necessary is easily understood if we look at this problem as the analogue problem of tiling a sphere with pentagons and hexagons and we take into account the topological properties of the sphere (Euler theorem, see, e.g., \cite{Iorio:2006ur}). The precise number of hexamers will not be fixed by this argument and needs further assumptions that we shall discuss in the next paragraph. The argument goes like this: Suppose that the ends of proteins only join three at the time. If $N_p$ is the number of $p$-mers used to tile a sphere of unit radius, i.e. $N_5$ pentamers and $N_6$ hexamers, the resulting polyhedron $P$ has $V_P = 1/3 \sum_N N_p p$ vertices, $E_P = 1/2 \sum_N N_p p$ edges, and $F_P = \sum_N N_p$ faces,
giving for the Euler characteristic $\chi = V_P - E_P + F_P$, the following expression
\begin{equation}\label{ngons}
  \sum_N (6 - p) N_p = 6 \chi = 12 \;,
\end{equation}
since for a sphere $\chi = 2$. Explicitly Eq.(\ref{ngons}) reads
\begin{equation}\label{sphere}
    (6 - 5) \;  N_5 + (6 - 6) \; N_6 = 12
\end{equation}
hence to tile a sphere $N_5 = 12$ is required, but $N_6$ can be arbitrary. As said, for virus capsids $N_6$ is not arbitrary but must be a specific number that we shall soon obtain. For the mathematical problem of the tiling of the sphere one might also imagine to use {\it heptagons}. In that case the Euler formula (\ref{ngons}) gives
\begin{equation}\label{heptagons}
N_5 - N_7 = 12 \;.
\end{equation}
Thus, starting from the tiling of the sphere with exactly 12 pentagons (and an arbitrary number of hexagons) one can add {\it pairs} pentagon-heptagon, but not a pentagon or a heptagon separately. Note that at this point this is only a mathematical consideration and its relevance for virus structure is all to be proved.

\begin{figure}
\centering
\includegraphics[height=.2\textheight]{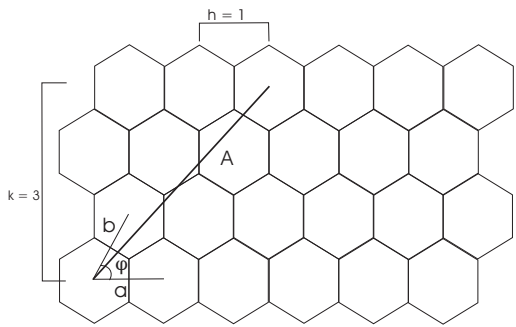}
  \caption{Planar hexagonal lattice of identical rigid proteins. The vector $\vec{A} = h \vec{a} + k \vec{b}$ corresponds to $h = 1$ and
  $k = 3$.}
  \label{fighex}
\end{figure}

The geometric interpretation of the Euler formula (\ref{ngons}) is that a sphere of unit radius has curvature $R_{\rm sphere} = + 1$ and each polygon contributes to this curvature with $R_p = (6 - p) / 12$: a hexagon with $R_6 = 0$, a pentagon with $R_5 = + 1/12$, a heptagon with $R_7 = - 1/12$. This can be understood by constructing hexagons, pentagons and heptagons out of equilateral triangles of paper. A hexagon is obtained by joining together 6 triangles: they all stay in a plane. Take one triangle out and join what is left to make a pentagon and the resulting figure will bend outwards, while adding one triangle to the hexagon to make a heptagon results into an inward bending. This also tells us that a certain amount of bending energy $E_b$ is necessary to convert a hexagon into a pentagon or into a heptagon. How big is $E_b$ depends on the elastic properties of the material used. Let us now describe in more detail the CK construction.

\begin{figure}
\centering
  \includegraphics[height=.2\textheight]{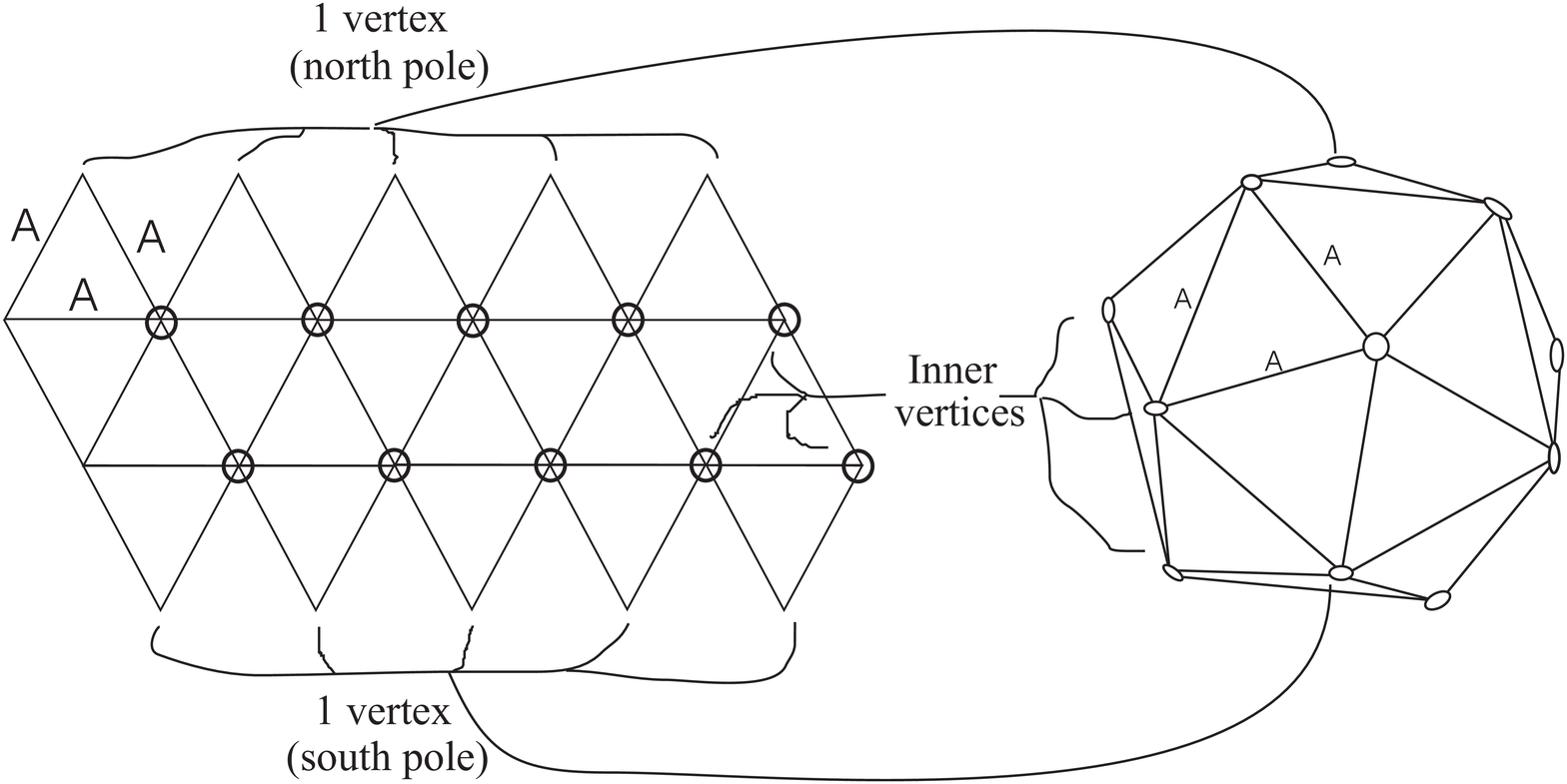}
  \caption{The equilateral triangles template and the icosadeltahedron. The 10 circled points on the planar template correspond to the 10 inner vertices of the solid, while all the outer vertices of the 5 upper triangles correspond to the north pole vertex of the solid and all the outer vertices of the 5 lower triangles correspond to the south pole vertex. At these locations the hexamers turn into pentamers. Each triangular face of the icosadeltahedron is made of $[T/2]$ hexamers (6 for the example of Fig.\ref{fighex}).}
  \label{bigtriangles}
\end{figure}

\paragraph{The CK construction} Suppose that the proteins are arranged on a plane to form the hexagonal lattice of Fig.\ref{fighex}.
Each side of the lattice represents a real protein. The basic vectors $\vec{a}$ and $\vec{b}$, with $|\vec{a}| = |\vec{b}|$, join the center of the hexagon taken as the origin of the lattice with the centers of the nearest hexagons as in figure. The angle is evidently $\varphi = 60^o$.  The 3-dimensional polyhedron these proteins will eventually form is obtained by imagining the 20 equilateral triangles with side $|\vec{A}| = A$ - where $\vec{A} = h \vec{a} + k \vec{b}$, and $h,k = 0, 1, 2,...$ - represented in Fig.\ref{bigtriangles} folded to obtain the icosahedron, the platonic solid with 12 vertices, 20 faces and 30 edges. Each triangle face of the icosahedron, contains a fixed number of hexamers that are the real proteins. At each of the 12 vertices the hexamers must turn into pentamers for the topological and geometrical reasons described above. Say $|\vec{a}| = a$, then one has $A^2 = a^2 (h^2 + k^2 + 2 h k \cos \varphi) = a^2 (h^2 + k^2 + hk) \equiv a^2 T(h,k)$, with $T(h,k) = 1,3,4,7,...$.
Being the area of the triangle given by $\alpha_A = (\sqrt{3}/4) a^2 T(h,k)$ and the area of one hexagon $\alpha_6 = (\sqrt{3}/2) a^2$, the number of hexagons per triangle is $n_6 = \alpha_A / \alpha_6 = [T/2]$. The total number of subunits is obtained by counting the total number of hexagons used for the {\it planar} lattice of Fig.\ref{fighex}, which is $N_6 = 20 (T/2) = 10 T$, then multiplying by 6 (the number of edges of the hexagon):
$N_{\rm proteins} = 60 T$. On the real 3-dimensional solid (that one one might think of obtaining by folding the planar template) the $60 T$ proteins are arranged as: i) $60$ form 12 pentamers; ii) $60 (T - 1)$ form $10 (T -1)$ hexamers, for a total number of morphological units of $N = 10 T + 2$. The figures obtained are icosadeltahedrons characterized by the pair of integers $(h,k)$ which not only are related to the total number of proteins, but also give the ``chirality'' of the polyhedron. Viruses belonging to this class follow these prescriptions with great accuracy and they are classified according to the values of $T$ (see Table \ref{ckclassif} for some examples and Ref.\cite{viper} for an exhaustive database on icosahedral virus structures).
\begin{table}
\begin{tabular}{lrrrr}
\hline
  & $N_{\rm Proteins}$
  & $T$  \\
\hline
Feline Panleukopenia Virus & 60 & 1 \\
Human Hepatitis B & 240 & 4 \\
Infectious Bursal Disease Virus (IBDV) & 780 & 13 \\ {\it General} & 60 T & $h^2 + k^2 + hk$ \\
\hline
\end{tabular}
\caption{Examples of viruses that follow the CK classification taken from Ref. \cite{viper}.}
\label{ckclassif}
\end{table}
Recently there have been various attempts to generalize the CK model to include also certain complex viruses. One of those attempts is the model proposed in Ref.\cite{nguyen} - based on the continuum elastic theory of large spherical viruses of Ref.\cite{lmn} - where the authors address the problem of understanding the formation of spherocylindrical and conical virus capsids. Later we shall show that, if a change in the texture of the arrangement of proteins (scar) takes place, those and many more shapes could be obtained.

\section{Lessons from the Thomson Problem}

\paragraph{Thomson Problem} Let us now turn our attention to a different but geometrically related physical set-up from which we would like to gain some insights for the generalization of the CK construction we are looking for: the Thomson problem \cite{thomson}. It consists of determining the minimum energy configuration for a collection of electrons constrained to move on the surface of a sphere and interacting via the Coulomb potential. This old (and largely unsolved) problem has many generalizations for more general repulsive potentials as well as for topological defects rather than unit electric charges \cite{bnt}, \cite{bnt2}. The fact that the two problems (virus capsids construction and equilibrium configurations for charges on a sphere) are intimately related can be seen from the numerical results for the Thomson problem that have been obtained over the years. In Ref. \cite{altschuler} the authors proposed as solution of the problem an arrangement of $N$ electrons on the sphere into a triangular lattice where each electron has 6 nearest neighbors sitting at the vertices of an hexagon, with the exception of 12 locations where the nearest neighbors are only 5 sitting at the vertices of a pentagon and $N = 10 T + 2$, with $T = h^2 + k^2 + hk$: that is the icosadeltahedron. Note that in this case the electrons are constrained to be on the surface of the sphere, e.g. imagining the sphere as a metal, while the proteins have not such constraint. Furthermore, the polygons here are ``imaginary'', in the sense that only the vertices are real particles, while the edges are not.

\paragraph{Scars (and Pentagonal Buttons)} Further studies \cite{perez-garrido} have shown that, even for $N = 10 T + 2$ electrons, when $T$ is large enough (of the order of $10^2$), configurations which differ from the icosadeltahedron have lower energy than the corresponding icosadeltahedron. That is, when near one of the 12 pentagons two hexagons (let us call this a 5-6-6 structure) are replaced by a pair heptagon-pentagon (let us call this a 5-7-5 structure) to form a linear pattern called {\it scar}, the energy is lower than that of a configuration of 12 pentagons and all the rest hexagons. This indeed happens in numerical simulations for higher and higher number of electrons, where the scars become longer (e.g. 5-7-5-7-5, etc.), always respect the topological/geometrical constraint of Eq. (\ref{heptagons}), can spiralize or might even form exotic patterns like two nested pentagonal structures with five pentagons placed at the vertices of the outer pentagonal structure, five heptagons at the vertices of the inner pentagonal structure, and a pentagon in the common center (the vertex of the icosadeltahedron) (see, e.g., \cite{bnt} and references therein). The latter patterns are called {\it pentagonal buttons} and an explanation of their topological origin can be found in Ref.\cite{Iorio:2006ur}. Apparently, even more complicated structures can appear in numerical simulations \cite{bnt}. Scars have been experimentally found to be formed in spherical crystals of mutually repelling polystyrene beads self-assembled on water droplets in oil \cite{realscars}. The repulsive potential there is not the Coulomb potential, hence that is a particular instance of the generalized Thomson problem. These experimental findings confirm that, at least in the case of scars, things go along the lines of the above outlined analysis.

The lesson we learn from the Thomson problem is that under certain conditions it is energetically favorable to convert a pair 6-6, with zero total and local curvature ($0 = 0 + 0$) and zero bending energy, into a pair 5-7, again with zero total curvature but with nonzero local curvature ($0 = +1/12 - 1/12$) hence with nonzero bending energy given by $2 E_b$, where $E_b$ is necessary to convert a 6 into a 5 or into a 7.

\section{Scars and Virus Structure}

\paragraph{Our Conjecture} What we propose here is that, due to the interaction with the environment (and/or with the genetic material), the formation of scars of pentamers and heptamers can take place in virus capsids during the process of assembly of the proteins. The way we believe this happens is as follows: i) At first the proteins assemble to make an icosadeltahedron following the CK prescription. ii) At an intermediate stage, due to the interaction with the environment they form scars near the location of one or more of the 12 pentamers at the vertices of the icosadeltahedron. This interaction is necessary because the needed extra bending energy ($2 E_b$ in the case of the formation of what we might call a ``simple'' scar: 5-7-5) can only come from the environment. iii) Eventually, the capsids change shape, from spherical to non-spherical via the release of the bending energy into stretching energy at the location of the scar with the consequent ``annihilation'' of the 5-7 pair into a 6-6 pair. The resulting capsid has the usual morphological units, pentamers and hexamers, but not the spherical shape. Thus it is to be expected that in real viruses scars should not be visible in the final stage but they should drive a change in shape from spherical to non-spherical. It is plausible, though, that i) in experiments where artificial virus capsids are synthesized, scars could be actually seen at an intermediate stage of the assembly when the ``would-be-capsid'' is frozen at a suitable point in time; ii) not all scars are annihilated, hence some of them could be visible on the final capsid. Note that in the presence of scars, the total number of proteins needed is the same as for the icosadeltahedral case without scars (this follows from 6 + 6 = 5 + 7) while the number and type of morphological units changes (for one simple scar: 13 pentamers, 1 heptamer, $10 T - 12$ hexamers, etc.).

As said earlier, there is a strong interest today in trying to generalize the CK construction to include non-spherical viruses, important examples being the retroviruses that have spherical, spherocylindrical and conical capsids (see, e.g., Ref.\cite{HIV-1} and references therein). In the work of Ref.\cite{nguyen} the proposal that spontaneous curvature of the proteins in the capsids can drive a change in shape from spherical to spherocylindrical or conical shapes is extensively studied and the geometric construction of certain capsids (spherocilyndrical and conical) is carried out. The application of that approach to the case of retroviruses is then performed in Ref.\cite{HIV-1}, where the importance of the environment for the assembly of retrovirus capsids is pointed out. What we conjecture here is that the basis of these phenomena is the formation of scars. Our belief is based on the following observations: i) Scars appear in the geometrically related (generalized) Thomson problem; ii) Their formation/annihilation mechanism here seems to us a natural way to convert the energy given by the environment into bending energy (formation) and subsequently into stretching energy (annihilation); iii) This way a mechanism for producing a great variety of shapes (not only the spherocilyndrical or conical) is in place: the formation/annihilation of scars (simple or complex) in different locations on the intermediate icosadeltahedron (we suppose that this has to happen near the vertices).

Other authors have speculated that scars should occur in virus capsids \cite{realscars}. They expect scars to be formed only on large viruses and that means that they are expecting scars to be seen on the final capsid. This is an instance that we do not exclude but that is not essential for us as our main proposal is to ascribe the shape change to the scars formation/annihilation mechanism.

\begin{figure}
\centering
  \includegraphics[height=.2\textheight]{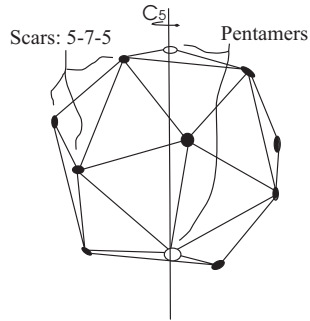}
  \caption{The intermediate spherical (icosadeltahedral) capsid with the $C_5$-symmetric distribution of simple scars.}
\label{s2}
\end{figure}

\paragraph{Variety of shapes} It is easy to convince oneself that indeed a great variety of shapes could be obtained via the scar formation/annihilation mechanism: At the site on the intermediate icosadeltahedron where the scar is formed and then annihilated the sphere gets stretched. The amount of stretching depends on the complexity of the scar\footnote{Complex scars might not be that rare as the same amount of energy is needed for the formation of, say, one next-to-simple scar (5-7-5-7-5) and two simple scars, i.e. $4 E_b$.}. The scars could be formed symmetrically (as we shall see in the next paragraph, for a particular symmetry of formation of scars we shall naturally obtain the spherocylindrical shape) or asymmetrically hence giving rise to regular or irregular shapes. Of all these very large number of shapes only a subset will describe real virus capsids because not all the shapes will be stable or energetically favored. A systematic study can be carried on using this method and case by case it could be seen whether it fits with the elastic properties of the virus capsids and with the constraints coming from the environment \cite{HIV-1}. What we shall do now is to construct, within our framework, one particular shape, the spherocylindrical. This will give us the opportunity to show how the method of construction works in a case that it is known to correspond to real virus capsids, like, e.g., certain bacteriophages.

\begin{figure}
  \centering
  \includegraphics[height=.2\textheight]{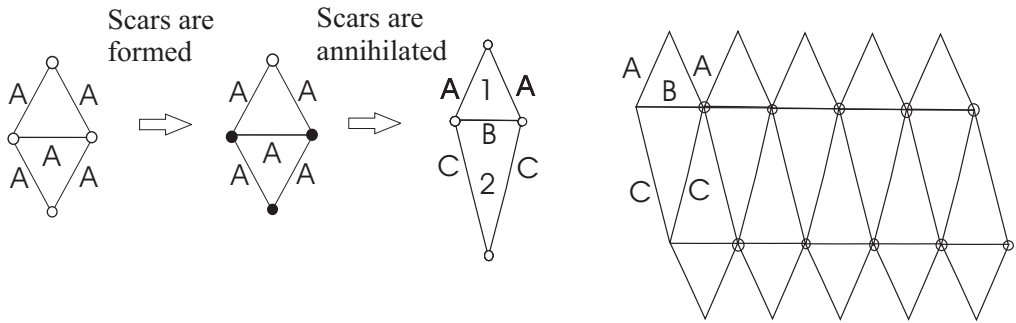}
  \caption{The generalized CK construction of the template driven by the scars formation-annihilation mechanism.}
\label{trianglesandtemplate}
\end{figure}

\begin{figure}
 \centering
  \includegraphics[height=.2\textheight]{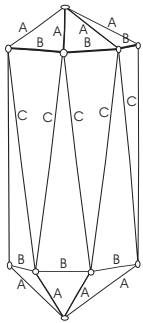}
  \caption{The spherocylindrical capsid.}
\label{spherocylinder}
\end{figure}

\paragraph{Spherocylindrical Capsids} Suppose that the intermediate icosadeltahedron is formed. We can then refer to the hexagonal lattice and to the template of Fig.\ref{fighex} and Fig.\ref{bigtriangles}. Let us imagine that the scars, e.g. all simple, are created only near the 10 inner vertices via a mechanism that respects the $C_5$ rotation symmetry\footnote{$C_n$ is the finite group of rotations of angles $2 \pi /n$, with $n= 1,2,3,...$. $C_5$ is one of the subgroups of the icosahedral group, the group of all possible symmetries of the icosahedron. Its relevance for the Thomson problem has been understood in \cite{Iorio:2006ur} where a mechanism of spontaneous symmetry breaking is seen as the responsible for some of the patterns found in numerical simulations. Here our introduction of the $C_5$ symmetry is motivated solely by the need to build up the spherocylinder.} around the north pole-south pole axis\footnote{Of course the axis is completely arbitrary as long as it encompasses two opposite vertices.}. In Fig.\ref{s2} the vertices where the scars are formed are indicated with $\bullet$, while the other two are indicated with $\circ$. Take a pair of the equilateral triangles of that template: any one from one of the outer layers of five triangles (e.g. the layer of triangles that correspond to the north pole) and the one from the inner layer that shares an edge with it. In Fig.\ref{trianglesandtemplate} of such pairs is shown and the different nature of the vertices is represented like in Fig.\ref{s2}. The scars are distributed in a way that is asymmetric with respect to the two triangles, hence the net effect of their formation/annihilation mechanism will deform them differently. Depending on the actual orientation of the scar around the given vertex the deformation will be different. To obtain the spherocylindrical capsid the three scars should make the lower triangle thinner and longer (they stretch the area and make it bigger) and this has the effect of shrinking the upper triangle by making the common edge shorter. Due to the symmetry of the location of scars the two edges of the new lower triangle have to be the same. If this mechanism takes place in the same fashion for all the ten pairs\footnote{Five north pole triangles paired with their common-edge inner triangles and five south pole triangles paired with their common-edge inner triangles.} of triangles of the template of Fig.\ref{fighex} the resulting new template is the one given in Fig.\ref{trianglesandtemplate}.
We require that this mechanism is area preserving, i.e. that the total number of proteins needed is the same as the one needed for the icosadeltahedron, they are only rearranged. This is obtained by requiring that $2 \alpha_A = \alpha_1 + \alpha_2$, where $\alpha_1$ is the area of the upper new triangle and $\alpha_2$ the area of the lower new triangle in Fig.\ref{trianglesandtemplate}. This means that the three quantities must be related as
\begin{equation}
\sqrt{3} A^2 = B \left(\sqrt{A^2 - \frac{1}{4} B^2} + \sqrt{{C}^2 - \frac{1}{4} B^2} \right) \;,
\end{equation}
with $B < A$ and $C > A$. Recall that, for $a=1$, $A^2 = T = h^2 + k^2 + h k$, hence the final capsid, obtained by folding the new template of Fig.\ref{trianglesandtemplate} (see Fig.\ref{spherocylinder}), will have (12 pentamers and) the $10 (T - 1)$ hexamers distributed differently with respect to the intermediate icosadeltahedron.

Notice that this spherocylinder is slightly different from the one obtained in \cite{nguyen} as the upper and lower half-icosadeltahedrons are not obtained by folding five equilateral triangles but five isosceles triangles (in this sense they are no longer proper half-icosadeltahedrons but a deformation of them). This is an instance that could be experimentally tested.

From this construction it is clear that a variety of shapes could be obtained this way. For instance, if the orientation of the scars in the previous setting is such that $C$ shrinks, hence $B$ becomes longer, then a disk-like shape is obtained. Let us stress here again that for this to correspond to real virus capsids one needs more detailed information on the elastic properties of the proteins and of the interaction with the environment.

\section{Conclusions}

In this paper we propose a mechanism of formation and subsequent annihilation of scars of pentamers-heptamers at an intermediate stage of the assembly of the virus capsid as the responsible for a great variety of non-spherical virus shapes. Our conjecture is based on the fact that scars are found in the (generalized) Thomson problem, in experiments and in numerical simulations, and on the observation that this mechanism would give a simple and plausible explanation of how the energy provided by the environment is converted into a change of capsid's shape. The conjecture can be tested, for instance, in experiments where artificial capsids are synthesized. Scars should appear on what we called here the intermediate icosadeltahedron, then should drive the change in shape. Capsids that could perhaps be used to this end are those relative to viruses that are known to have non-spherical final shape but still pentamers and hexamers as morphological units, like for instance certain bacteriophages. This conjecture, if experimentally confirmed,  would extend the classic Caspar and Klug construction for icosahedral viruses to include viruses that still have pentamers and hexamers as morphological units but no longer are icosadeltahedrons.

Let us conclude by making the remark that a better understanding of the way virus capsids are formed might suggest ways of destroying a virus by, for
example, making the capsid unstable.

\section*{Acknowledgments}
A.I. thanks Paul Voorheis of Trinity College Dublin, Daniel Grumiller of MIT Boston, for enjoyable discussions and for providing some difficult-to-find references and Arianna Calistri of the University of Padua for advice with virology. S.S. acknowledges the kind hospitality of the Institute of Particle and Nuclear Physics of Charles University Prague.



\bibliographystyle{aipprocl} 





\end{document}